\documentclass[%
preprint,
amsmath,amssymb,
aps,
]{revtex4-2}

\usepackage{graphicx}
\usepackage{dcolumn}
\usepackage{bm}

\begin{document}

\preprint{APS/123-QED}

\title{Phase-Amplitude Representation of Continuum States}

\author{Daniel Hadush}
\email{daniel.hadush@famu.edu}
\author{Charles Weatherford}
\email{charles.weatherford@famu.edu}
\affiliation{Physics Department, Florida A\&M University, Tallahassee, FL, USA}

\date{\today}

\begin{abstract}
  A numerical method of solving the one-dimensional Schr\"odinger equation for
  the regular and irregular continuum states using the phase-amplitude
  representation is presented. Our solution acquires the correct Dirac-delta
  normalization by wisely enforcing the amplitude and phase boundary values. Our
  numerical test involving point-wise relative errors with the known Coulomb
  functions shows that the present method approximates both the regular and
  irregular wavefunctions with similar, excellent accuracy. This is done by
  using new basis polynomials that, among other advantages, can elegantly
  enforce the derivative continuity of any order. The current phase-amplitude
  method is implemented here to study the continuum states of Coulomb-screened
  potentials. We discovered that, during the parametric transition from a
  Hydrogen atom to the Yukawa potential, the electronic density at the origin
  exhibits surprising oscillation~-~a phenomenon apparently unique to the
  continuum states.
\end{abstract}
\keywords{Phase amplitude method, electronic density, Yukawa potential}
\maketitle
%
%
\section{Introduction}%
\label{sec:introduction}

Wavefunctions of an electron in a continuum state oscillate like the
trigonometric sine and cosine functions as the effect of the potential fades out
at infinity. This behavior contrasts with their bound state counterparts, which
decay exponentially after a few oscillations near the origin. This inherent
oscillation makes the numerical computation of continuum states time-consuming
and prone to errors. They are also not square integrable, meaning numerical
integration cannot determine their normalization. However, it has long been
established that both of these issues can be efficiently addressed if one were
to instead solve for the amplitude and the phase of the continuum states, both
of which are universally simple, monotonic functions~\cite{PhysRev.35.863,
  PhysRev.52.1123}. The amplitude can also be appropriately written so that it
flattens into a constant at infinity (away from the potential effects). This
constant can be used to set the normalization.

However, the governing equations for the phase-amplitude representation are
nonlinear and numerically challenging, given that their solutions are
deceptively simple. In their attempt to solve these equations, researchers are mostly faced with a treacherous effort involving several trials to exhaust all reasonable options. In addition to
solving the differential equation(DE), the amplitude must have proper
normalization and the phase proper synchronization to be valid. Despite the
numerous literature on the topic, clear evidence has not yet been reported that
specifies a successful numerical computation~\cite{RAWITSCHER2016138,
  PhysRevE.69.035402, PhysRevA.97.022701}.

The path toward the solution presented in this work is the best we could achieve
after exhausting several options, and its outcome is portrayed by the point-wise
relative errors for the Coulomb functions, one of the rare systems with known
analytic solutions.

Electronic density at the origin of the Coulomb-screened potentials is also
studied in this work for the first time in light of the continuum states. The
bound state counterparts have been reported by the authors
recently~\cite{PhysRevE.108.045301}, and this work can be considered a
continuation of that study. The discussion begins with the tools
that will be used in the numerical computation, and then a schematic derivation
of the working DEs and detailed steps on how to solve them will be
given. Discussions on numerical examples, plots, and tables will also be
presented, followed by conclusions.
%
%
%

\section{Translation Operator}%
\label{sec:translation-operator}

The translation operator ${\operatorname{e}}^{\mathbf{x} \, . \,
  \nabla_{\mathbf{a}}}$ in linear algebra translates a function $f$ of $m$
variables from point $\mathbf{a}$ to point $\mathbf{a} +
\mathbf{x}$~\cite{jordan1969linear}.
\begin{equation}
  \label{eq:1}
   f(\mathbf{a} + \mathbf{x}) = {\operatorname{e}}^{\mathbf{x} \, . \,
     \nabla_{\mathbf{a}}} f(\mathbf{a}) = \sum_{n = 0}^{\infty}
   \frac{{\left(\mathbf{x} \, . \, \nabla_{\mathbf{a}} \right)}^n}{n!}
   f(\mathbf{a})
\end{equation}
\noindent We have used a boldface font for an m-dimensional vector $\mathbf{x} =
(x_1, x_2, \ldots, x_m)$. As shown above, the definition is also equivalent to
the Taylor's expansion of the function $f(\mathbf{a} + \mathbf{x})$ about a
fixed point $\mathbf{a}$. Expanding the dot product of the differential operator
in the Cartesian coordinate system as $\left(\mathbf{x} \, . \,
  \nabla_{\mathbf{a}} \right) = \sum_{i = 1}^m x_i \left( d / d a_i \right)$,
using the multinomial expansion of its $n^{\rm{th}}$ power, and substituting
into Eq.~(\ref{eq:1}), one can write
\begin{equation}
  \label{eq:2}
  f(\mathbf{a} + \mathbf{x}) = \sum_{n = 0}^{\infty} \sum \left[
    \left(\frac{x_1^{n_1}}{n_{1}!} \frac{d^{n_1}}{da_1^{n_1}} \right)
    \left(\frac{x_2^{n_2}}{n_{2}!} \frac{d^{n_2}}{da_2^{n_2}} \right) \ldots
    \left(\frac{x_m^{n_m}}{n_{m}!} \frac{d^{n_m}}{da_m^{n_m}} \right)  \right]
  f(\mathbf{a})
\end{equation}
\noindent where the inner summation is over all different combinations of
non-negative integers $n_1, n_2, \ldots, n_m$ with $\sum_{i = 1}^m n_i =
n$. Hence, new basis polynomial functions, termed by us `Taylor' basis (TB)
polynomial, defined in Appendix~\ref{sec:tayl-basis-polyn}, naturally emerge
from Eq.~(\ref{eq:2}). Their implementation here will be in the numerical
approximation of functions locally, in the finite region near point
$\mathbf{a}$, by writing the finite form of Eq.~(\ref{eq:2}) in terms of the TB
as shown below.
\begin{equation}
  \label{eq:3}
  f(\mathbf{a} + \mathbf{x}) = \sum_{n = 0}^{N-1} \sum \left[ T_{n_1}(x_1)
    T_{n_2}(x_2) \ldots T_{n_m}(x_m) \right] C_{n_1 n_2 \ldots n_m}, \quad N =
  1, 2, \ldots
\end{equation}
\noindent This expansion’s principal significance is that the coefficients $C$
are numerically the higher order derivatives of $f$ at $\mathbf{a}$, i.e.,
$C_{n_1 n_2 \ldots n_m} = \left[ d^{(n_1 n_2 \ldots n_m)} / da^{(n_1 n_2 \ldots
    n_m)} \right] f(\mathbf{a})$. The parametric dependence of the $C$'s on
$\mathbf{a}$ is not explicitly shown above to avoid clutter. One can also drop
$\mathbf{a}$ from the equation entirely once the variable $\mathbf{x}$'s
locality to the fixed point $\mathbf{a}$ is implicitly understood.

Thus, the TBs are very convenient for implementing boundary conditions in
applications based on finite elements or volume. This precise definition of the
coefficients $C$ also implies that their numerical value is closely related to,
or is consistent with, the underlying physical model. Moreover, the association
of the rapidly growing factorials with the TB leaves the $C$'s with a rather
uniform and intuitive numerical behavior, even for a high-order expansion. They
are better suited for the collocation method, especially in contrast with the
ordinary power function $x^n$, as it is well known that the latter leads to a
Vandermonde matrix, which is an ill-conditioned matrix~\cite{golub13,
  Press:2007:NRE:1403886}.

The outer summation on $n$ above specifies an order of expansion equivalently
meaningful in any dimension. It is a means by which we can systematically
increment the order of Taylor’s series representation of the translation
operator by including the batch of terms according to the inner sum in
Eq.~(\ref{eq:3}), whose count for each $n$ is known to be $\binom{n + m - 1}{m -
  1}$~\cite{Olver:2010:NHM:1830479}. Equivalently, the total number of terms in
the expansion for a given $N$ is $\frac{N}{m} \binom{N + m - 1}{m - 1}$. The
term in braces in the last two expressions is the binomial coefficient.

\subsection{One-dimensional Expansion}
\label{sec:one-dimens-expans}

In this section, we will describe a spectral element method on how the TB can be
used to solve a homogeneous differential operator of type $\hat{S} (x) f(x) = 0$
for a local variable $x$ corresponding to a point inside a segment $[a, b]$ such
that $0 \le x \le b - a$. The one-dimensional case will be written for the
current purpose as
\begin{equation}
  \label{eq:4}
  f(x) = \sum_{n = 0}^{\nu - 1} T_{n}(x) C_{n} + \sum_{n = \nu}^{N - 1}
  T_{n}(x) C_{n}.
\end{equation}
The first $\nu$ of the coefficients $C$ are assumed to be known from the
boundary condition at point $a$, i.e, $C_n = f^{(n)}(a)$. Combined with the
collocation method, this typically leads to a set of simultaneous equations
shown element-wise below.
\begin{equation}
  \label{eq:5}
  \sum_{n = \nu}^{N - 1} S_{m n} C_{n} = F_m, \qquad m = 0, 1, \ldots, N - \nu - 1
\end{equation}
\noindent where
\begin{equation}
  \label{eq:6}
  S_{m n} = {\left[ \hat{S} (x) T_{n}(x) \right]}_{x = x_m}, \quad {\rm{and}}
  \quad F_{m} = -\sum_{n = 0}^{\nu - 1} S_{m n} f^{(n)}(a).
\end{equation}
The collocation point $x_m$ is the ${(m + 1)}^{\rm{th}}$ node of the Legendre
polynomial of order $N - \nu$ that is properly mapped onto $[0, b - a]$. The
value of $\nu$ depends on the order of the DE in question. For a second-order
DE, for instance, $\nu = 2$ because we must enforce the continuity of the
function and its first derivative across the boundary of consecutive finite
elements. The notable exception is the very first element, where it suffices to
set only $f(a)$ since any function is of zeroth order from the discontinuous
direction of an endpoint.
\section{Phase-Amplitude Representation}%
\label{sec:phase-ampl-repr}

In this work, we shall use an alternate representation of the regular and
irregular continuum wavefunctions, to be denoted here as $S_l(r)$ and $C_l(r)$,
respectively, with the following phase-amplitude \emph{ansatz}
\begin{align}
  \label{eq:7}
  S_l(r)
  &= A_l(r) \sin\left[ \Phi_l(r) \right] \\
  \label{eq:8}
  C_l(r)
  &= A_l(r) \cos\left[\Phi_l(r) \right],
\end{align}
\noindent with inverse relations,
\begin{align}
  \label{eq:9}
  A_l(r)
  &= \sqrt{S_l^2(r) + C_l^2(r)} \\
  \Phi_l(r)
  \label{eq:10}
  &= \tan^{-1} \left[\frac{S_l(r)}{C_l(r)} \right]
\end{align}
\noindent where $A$~\&~$\Phi$ are the amplitude and phase functions,
respectively. An immediate consequence of the above representation is obtained
by taking a derivative of Eq.~(\ref{eq:10})
\begin{equation}
  \label{eq:11}
  \frac{d}{d r} \Phi_l(r) = \frac{W_l(r)}{A_l^2(r)}
\end{equation}
\noindent where we have recognized the Wronskian as $W_l(r) = C_l(r) \dot{S}_l(r) -
S_l(r) \dot{C}_l(r)$. The overhead dot represents the derivative with respect to
the argument. For the radial form of the Schr\"odinger equation, which is
homogeneous second-order DE, $W$ is solely dependent on the coefficient of the
first-order term. This makes Eq.~(\ref{eq:10}) fundamental because it remains
true even without stating the potential.

We begin discussing a quantum mechanical system with the reduced form of
the radial equation for an electron in a continuum state under a potential
$V(r)$
\begin{equation}
  \label{eq:12}
  \ddot{\psi}_l(r) + Q_l(r) \psi_l(r) = 0
\end{equation}
\noindent where
\begin{equation}
  \label{eq:13}
  Q_l(r) = k^2 - \frac{l(l + 1)}{r^2} - 2 V(r).
\end{equation}
Atomic units will be used throughout. The standard wavefunction will be denoted
by the capital $\Psi_l(r)$, where its related to its reduced form shown above as
$\psi_l(r) = r \Psi_l(r)$. The total energy of the system $E$ is written in
terms of the momentum number $k = \sqrt{2 E}$. The wavefunction $\psi$
represents either of the regular $S$ or irregular $C$ functions. One of the
advantages of using the reduced form is, from what is commonly referred to as
Abel's identity~\cite{hassani2013mathematical}, the resulting $W$ takes its
simplest form: a constant. Moreover, at infinity, the amplitude asymptotically
goes to a constant while the two continuum solutions it envelopes behave like
the trigonometric sine and cosine functions. For wavefunctions normalized in the
energy scale~\cite{BARSHALOM199621, Fischer_1997}
\begin{equation}
  \label{eq:14}
  \int_0^\infty \psi_l(E, r) \psi_l(E', r) dr = \delta (E - E'),
\end{equation}
their correct values are $W_l(r) \equiv W = 2 / \pi$ and $\lim_{r \to \infty}A
_l(r) = \sqrt{2 / (k \pi)}$. These two values are fundamental to calculating
appropriately normalized solutions. Momentum normalized $\delta (k - k')$
solutions can be obtained by multiplying with $\sqrt{k}$~\cite{Bethe_2014}. In
all of the examples in this article, $k = 1$ is used to avoid ambiguity.

To find the DE for the amplitude, we differentiate Eq.~(\ref{eq:7}) twice,
substitute the regular solutions into Eq.~(\ref{eq:12}), and group the resulting
terms into factors of $\sin \Phi$ and $\cos \Phi$. Demanding that all the terms
that are factors of the cosine vanish gives $A \ddot{\Phi} + 2 \dot{A}
\dot{\Phi} = 0$, for which Eq.~(\ref{eq:11}) with a constant $W$ is a
solution. Doing the same with the factors of the remaining sine terms and using
Eq.~(\ref{eq:11}) gives the following nonlinear DE, commonly known as Milne
DE~\cite{PhysRev.35.863}.\@
\begin{equation}
  \label{eq:15}
  \ddot{A}_l(r) + Q_l(r) A_l(r) = \frac{W^2}{A_l^3(r)}
\end{equation}
In most literature, the numerator on the right-hand side has been written as
some constant, mostly $k^2$, or its numerical value~\cite{10.1063/1.4929399,
  BARSHALOM199621, PhysRevA.97.022701}. For
clarity, we have explicitly written out the quantity it represents, the
Wronkskian, instead of its actual value.

The solutions $A$ and $\Phi$ are considered valid only if they are appropriately
normalized and can attain both triangular legs $S$ and $C$ with similar
accuracy. The best way to enforce normalization will be to integrate in reverse
direction after an appropriate variable change. In this work, $\rho = 1 / r$,
where $0 \le \rho < \infty$ is implemented. Unfortunately, the resulting DE of
the amplitude in the new variable $\rho$ becomes even more nonlinear and,
according to our experience, difficult to solve in a stable manner. A possible
remedy is to solve instead for the amplitude squared $Y$ or $A_l(r) =
\sqrt{Y_l(r)}$. It is easy to show that a direct substitution into
Eq.~(\ref{eq:15}) gives the following nonlinear DE.\@
\begin{equation}
  \label{eq:16}
  \ddot{Y}_l(r) + 2 Q_l(r) Y_l(r) = \frac{1}{2 Y_l(r)} \left[ \dot{Y}^2_l(r) +
    {\left( 2 W \right)}^2 \right]
\end{equation}
\noindent It has also been reported that $Y$ obeys the following linear DE,
which is third order and requires knowledge of the derivative of the potential
$V$~\cite{10.1063/1.4929399, PhysRevA.97.022701}.
\begin{equation}
  \label{eq:17}
  \dddot{Y}_l(r) + 4 Q_l(r) \dot{Y}_l(r)  + 2 \dot{Q}_l(r) Y_l(r) = 0
\end{equation}
\noindent The above two DEs are readily solvable in the $\rho$ coordinate. The outline of
the steps will be summarized below.

\subsection{Amplitude}%
\label{sec:amplitude}

After rewriting the above two DEs in terms of $\rho = 1 / r$ and dividing the
$\rho$-axis into finite elements, $Y$ can be integrated out for $\rho \ge 0$
using the method described in~\ref{sec:one-dimens-expans}. The integration must
stop at some $\rho = r_{\min}^{-1}$ because $Y$ (due to the irregular $C$) is
singular at $r = 0$. A prerequisite for an upper limit of $r_{\min}$ will be given
below shortly. For the first element, simply use $\nu = 1$ and fix
$C_0 = 2 / (k \pi)$, which takes care of the normalization. This is convenient
because the higher derivatives of $Y(r)$, which were all zero at $r = \infty$,
change non-trivially for $\rho = 0$ upon conversion. However, we do not need to
find out their value to start the integration here, which is advantageous over
other methods that do. This is true for ordinary DE solvers that are naturally
first-order and must cast higher order DEs into a matrix form. For the rest of
the following elements, $\nu = 2$ and $3$ shall be used for
Eq.~(\ref{eq:16})~\&~Eq.~(\ref{eq:17}), respectively. Since Eq.~(\ref{eq:16}) is
nonlinear, it must be solved iteratively where the nonlinear terms on the
right-hand side are recovered from the previous iteration. Convergence can be
expedited by seeding the coefficients $C$, as defined, with the respective
derivatives of the solution at the endpoint of the previous element, thus
setting the iteration off to a very good start.

\subsection{Phase}%
\label{sec:phase}

Phase $\Phi$ is a monotonic function that gains an amount of $\pi$ at every node
of $S$ (or $C$). Therefore, it is only intuitive to set, by convention, its
lowest value at the origin. In our case, $\Phi_l(r = 0) = 0$ since we are
considering the reduced form of the solution. After $Y$ is known, part of the
first element near $\rho = 0$ does not hold a valid information about $\Phi$
because the distance has artificially been made finite. The integration of
$\Phi$ must be done in the forward direction, as written in Eq.~(\ref{eq:11}),
which confronts us with the synchronization problem of determining the
integration constant. A reliable and versatile option is to exploit the fact
that given the above convention, $\Phi_l(r_{\pi}) = \pi$ can be used where
$r_{\pi}$ is the first root of the regular solutions $S$. Then the phase will
simply be
\begin{equation}
  \label{eq:18}
  \Phi_l(r) = \pi + W \int_{r_{\pi}}^r \frac{d x}{Y_l(x)} \rm{.}
\end{equation}
A regular solution with arbitrary normalization is readily solvable numerically
using, for instance, the recently developed first-order forms of the radial
equation~\cite{PhysRevE.108.045301} and some of the techniques discussed in this
article. The additional work this entails is rewarded by an accurate value of
$r_{\pi}$ thanks to the available, excellent root-finding algorithms such as the
Newton-Raphson method. Remember, we already need to resort to other methods if
we are interested in $S$ from $0$ to $r_{\min}$. This also implies a new upper
limit on the choice of $r_{\min}$, namely $r_{\min} < r_{\pi}$.

\subsection{Regular and Irregular Solutions}
\label{sec:regul-irreg-solut}

Once $Y$ and $\Phi$ are obtained, for $r \ge r_{\min}$, the regular and
irregular solutions can simply be calculated by dropping the two legs of the
hypotenuse $A$ according to Eq.~(\ref{eq:7})~\&~Eq.~(\ref{eq:8}),
respectively. When the phase is appropriately synchronized, its $\sin(\Phi)$ and
$\cos(\Phi)$ will be equally correct, such that $S$~\&~$C$ are obtained with
identical accuracy. We are now also in a position to get the normalization
constant for the regular solution calculated above to obtain $r_{\pi}$ and use
the resulting, properly normalized $S$ in the region that extends to the origin
($0 \le r \le r_{\pi}$). In this sense, how below $r_{\pi}$ one would like to
lower $r_{\min}$ is motivated entirely by the less common irregular solutions
$C$.

\section{Example: Screened Coulomb Potentials}%
\label{sec:yukawa-potential}

\begin{figure}
  \includegraphics[scale=1.0]{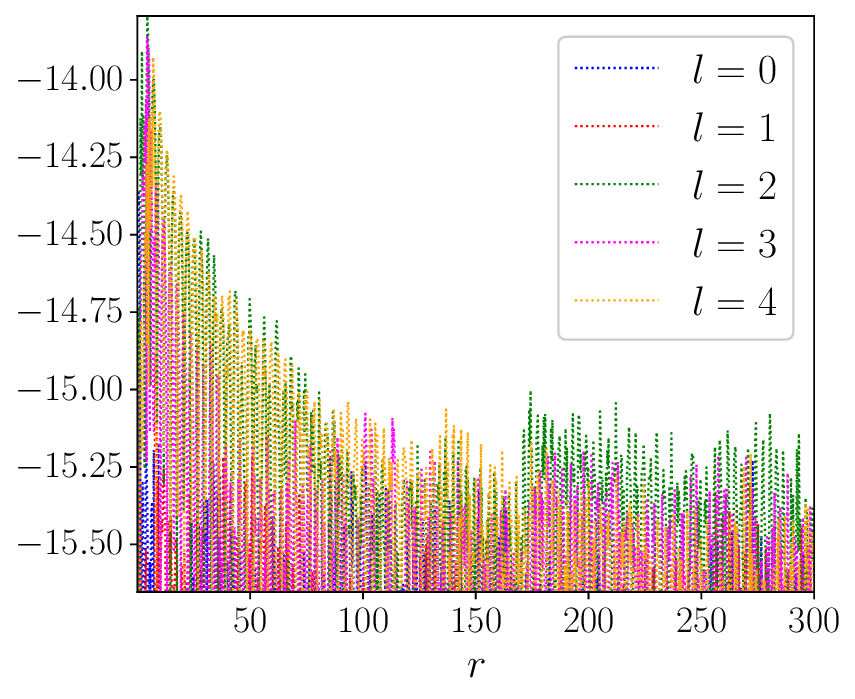}
  \caption{(Color online) $\log_{10}$ of relative error of $l \le 4$
    states of the regular Coulomb functions for $0.001 \le r \le 300$.}%
  \label{fig:fig1}
\end{figure}

\begin{figure}
  \includegraphics[scale=1.0]{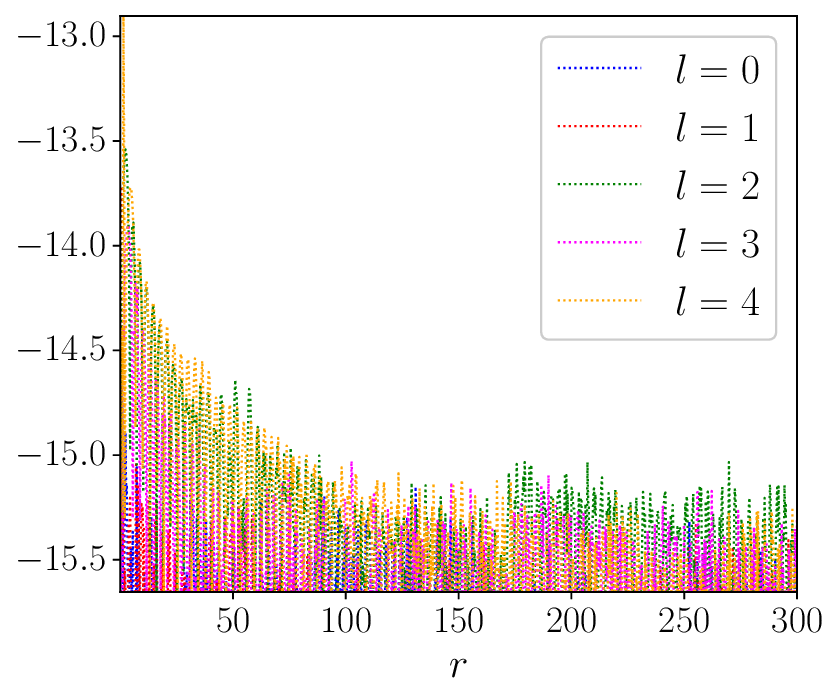}
  \caption{(Color online) $\log_{10}$ of relative error of $l \le 4$
    states of the irregular Coulomb functions for $0.001 \le r \le 300$ as
    calculated from Eq.~(\ref{eq:8}). The
    integration was done from left to right.}%
  \label{fig:fig2}
\end{figure}

As a check, we calculated the regular and irregular Coulomb functions $V(r) = -Z
/ r$, for which
analytic solutions are known and are commonly available in scientific
libraries~\cite{SEATON2002225, Olver:2010:NHM:1830479, contributors-gsl-gnu-2010}. For each finite element, the order of the TB polynomials $T$ in
Eq.~(\ref{eq:4}) is set as $N = \nu + 12$. For demonstration, $r_{\min} = 0.001$
is also used. The resulting relative errors of the states $s, p, \ldots$,
belonging to the first few angular quantum numbers are shown in
Figs.~\ref{fig:fig1}~\&~\ref{fig:fig2}.

Strictly speaking, the plots are produced from the linear Eq.~(\ref{eq:17}). Our
output from the non-linear Eq.~(\ref{eq:16}) (not shown here) is qualitatively
similar overall, with errors that are not worse than by an order of
magnitude. The traditional method is used to calculate the regular solutions in
the inner segment $r < r_{\pi}$ to depict a pragmatic implementation, which
explains the improved accuracy fashioned in Fig.~\ref{fig:fig1}. The two plots
are otherwise similar, characteristic of a well-synchronized phase. This can be
imposed as a sanity check on the $\Phi$ in similar applications. Reference
values are taken from~\cite{contributors-gsl-gnu-2010}.

\begin{figure}
  \includegraphics[scale=1.0]{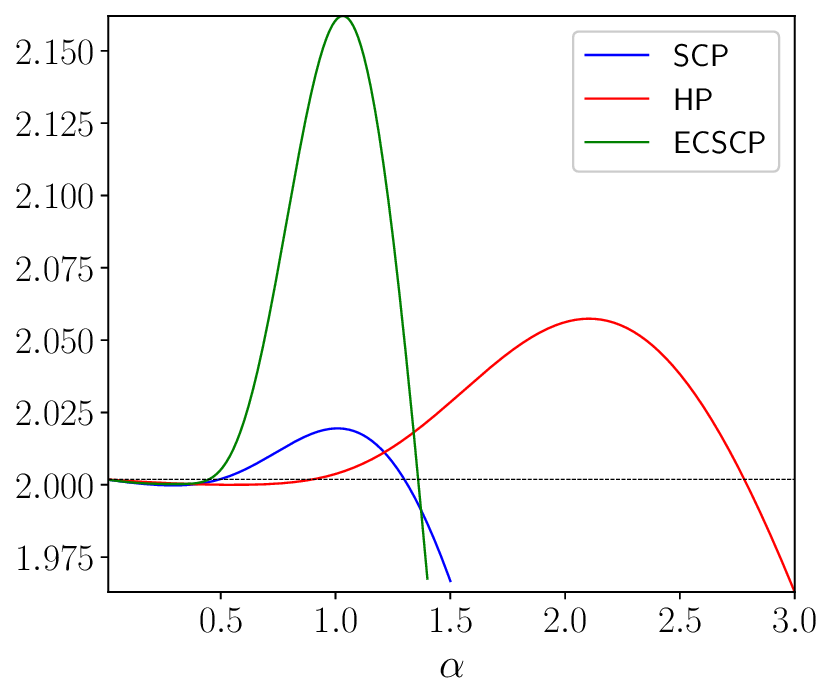}
  \caption{The normalized $1s$ state wavefunction at the origin
    versus $\alpha$ for the indicated Coulomb screened potentials. The dotted
    horizontal line is at $2.001\;870\;062\;315\;39$ for the limiting case
    $\alpha = 0$ (Hydrogen atom). The plots are for $\alpha \ge 0.01$ and
    $k = 1$.}%
  \label{fig:fig3}
\end{figure}

\begin{figure}
  \includegraphics[scale=1.0]{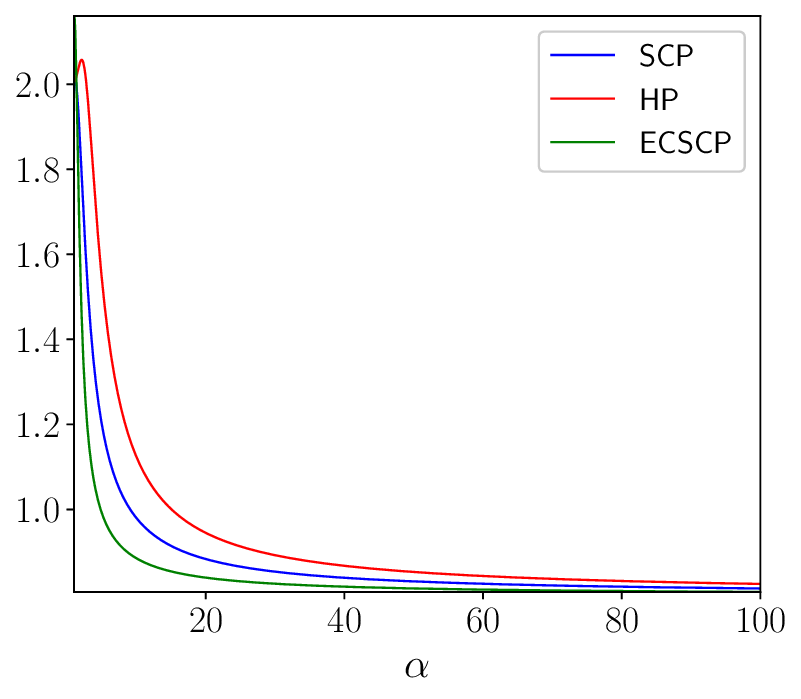}
  \caption{(Color online) A continuation of Fig.~\ref{fig:fig3} for $1 \le \alpha
    \le 100$.}%
  \label{fig:fig4}
\end{figure}

Now that we can calculate the normalized continuum states with
the demonstrated accuracy, we would like to report a subtle phenomenon in a
closely related system, that comes to light only with sound numeric
output. Specifically, the electronic density of the Yukawa potential and its
close relations, the bound state of which were studied by us recently(REf): the Yukawa
or static screened Coulomb potential (SCP), exponential cosine screened Coulomb
potential (ECSCP), and Hult\'{h}en potential (HP)
\begin{equation}
  \label{eq:19}
  V (r) =
  \begin{cases}
    -\frac{Z {\operatorname{e}}^{-\alpha r}}{r}
    & \mbox{for SCP,} \\
    \frac{{\operatorname{e}}^{-\alpha r}}{(1 + {\operatorname{e}}^{-\alpha
    r})} \frac{Z \alpha }{\tanh(-\alpha r / 2)}
    & \mbox{for HP,} \\
    -\frac{Z {\operatorname{e}}^{-\alpha r}}{r} \cos{\alpha r}
    & \mbox{for ECSCP.}
  \end{cases}
\end{equation}
Only $Z = 1$ is used. The screening parameter $\alpha \ge 0$ links all of the
above three systems to the hydrogen atom at $\alpha=0$. The bound states of the
above three systems, specifically, how the ground state energy and the
electronic density at the origin vary as $\alpha$ runs from zero to its critical
value, where a bound state can no longer be supported, have been studied
recently. In this article, a similar study is done with the electronic density
at the origin, here $\Psi_0^2(0)$, but for the continuum states. These are shown
in a separate plots in Figs.~\ref{fig:fig3}~\&~\ref{fig:fig4} to zoom in on the
peculiar oscillation of the density exhibited by small $\alpha$.
We expected this particular quantity to decrease monotonically, much like it
does in Fig.~\ref{fig:fig4}, or in a similar fashion to its bound
counterpart. Instead, we obtained the single oscillation shown in
Fig.~\ref{fig:fig3}. This phenomenon is interesting because it happens for small
$\alpha$, just as the system leaves the $H$ atom and not near the critical
value. It is not clear what causes its shape or why it does not repeat
elsewhere. For reference purposes, representative benchmark values $\Psi_0(0)$
for the three potentials, including their maximum values, are given in
Table~\ref{tab:table1}.

All of the plots in this article comprise 1025 uniform data points each.
%
%
%
\begin{table*}
  \caption{Electronic density at the origin, $\Psi_0^2(0)$, for the regular
    $1s$ continuum state and $k = 1$ of the Coulomb screened potentials. The
    underlined figures given in the lower half of the table show the coordinates
    for the peak points in Fig.~\ref{fig:fig3}}%
  \centering
  \begin{ruledtabular}
    \begin{tabular}{cccc}
      & \multicolumn{3}{c}{$\Psi_0(0)$} \\
      \cline{2-4}
      \textrm{$\alpha$}
      & SCP
      & HP
      & ECSCP \\
      \hline
      0.01
      & 2.001 753 468 852 04
      & 2.001 811 569 983 04
      & 2.001 754 351 507 03
      \\
      0.1
      & 2.000 812 154 698 96
      & 2.001 317 067 239 68
      & 2.000 899 143 538 91
      \\
      1
      & 2.019 489 044 554 80
      & 2.003 738 379 326 01
      & 2.160 510 127 337 48
      \\
      10
      & 0.979 691 498 590 326
      & 1.123 708 852 625 44
      & 0.885 015 629 968 403
      \\
      100
      & 0.814 064 055 523 049
      & 0.824 743 905 140 524
      & 0.805 927 073 382 779
      \\
      \hline
      \underline{1.008 625 496 800 67}
      & \underline{2.019 501 542 762 27}
      & 2.003 944 948 800 74
      & 2.161 241 023 924 67
      \\
      \underline{2.104 242 184 705 96}
      & 1.785 741 131 308 62
      & \underline{2.057 415 252 014 99}
      & 1.451 864 737 136 60
      \\
      \underline{1.031 378 946 011 15}
      & 2.019 411 732 469 42
      & 2.004 526 760 324 64
      & \underline{2.162 061 665 512 45}
    \end{tabular}
  \end{ruledtabular}%
  \label{tab:table1}
\end{table*}

\section{Conclusions}%
\label{sec:conclusions}

The viable path for solving continuum systems using the powerful phase amplitude
representation presented here should be sufficient for most applications in
electronic structure calculations and can serve as a template. Both of the
second and third order DEs for amplitude can be integrated in just a few steps
using the included algorithm based on the spectral method within the finite
elements. They enable us to arrive at regular and irregular solutions with
identical accuracy simultaneously.

TB polynomials are easy to understand and possess properties given in
Appendix~\ref{sec:tayl-basis-polyn} that are evidently convenient to
implement. The direct connection of the expansion coefficients with the
gradient operator makes them a powerful tool for enforcing derivative continuity
of arbitrary order across boundaries or for accurate initialization of iterative
schemes. Since they stem from the basic translation operator, their potential
utility in other avenues deserves more attention, especially in higher
dimensional applications.

What is also worth attention to is the oscillation phenomenon of the electronic
density at the origin presented, which is observable only with accurately
normalized solutions. It should be further understood and put into perspective
with other parameterized systems. The behavior of this simple physical quantity
affirms that a parametric transition between two physical regimes is not always
a smooth sail and perhaps not necessarily intuitive.
%
%
%
\begin{acknowledgments}
  D.H. and C.W. were partially supported by the Department of Energy, National
  Nuclear Security Administration, under Award Number (s) DE-0003984.
\end{acknowledgments}
\appendix

\section{Taylor Basis Polynomials}
\label{sec:tayl-basis-polyn}

For a real number $x$, the `Taylor' basis (TB) polynomial functions are defined
as
\begin{equation}
  \label{eq:A1}
  T_n(x) = \frac{x^n}{n!}, \quad n = 0, 1, 2, \ldots
\end{equation}
\noindent where $n!$ is the factorial function. They obey a simple two-term recurrence formula
\begin{equation}
  \label{eq:A2}
  T_{n}(x) = \frac{x}{n} T_{n - 1}(x), \qquad T_0(x) = 1.
\end{equation}
\noindent Differentiation and integration operations act as the ladder (lowering
and raising) operators in quantum mechanics as shown below.
\begin{equation}
  \label{eq:A3}
  \frac{d}{dx} T_{n}(x) = T_{n - 1}(x), \qquad \int_0^x T_n(y) \, d y =
  T_{n + 1}(x)
\end{equation}
\noindent Their product is another TB of higher order scaled by a binomial
coefficient.
\begin{equation}
  \label{eq:A4}
  T_{m}(x) \, T_{n}(x) = \binom{m + n}{n} T_{m + n}(x)
\end{equation}
\noindent Finally, their addition formula is given by
\begin{equation}
  \label{eq:A5}
  T_{n}(x + y) = \sum_{m = 0}^n T_{m}(x) \, T_{n - m}(y).
\end{equation}
\bibliography{FAMUAPStwoPhA}

\end{document}